\pdfoutput=1
\documentclass[10pt]{article}
\usepackage[utf8]{inputenc}
\usepackage[T1]{fontenc}
\usepackage{lmodern}
\usepackage[a4paper,margin=2.2cm]{geometry}
\usepackage{amsmath,amssymb,amsthm}
\usepackage{graphicx}
\usepackage{booktabs,tabularx}
\usepackage{algorithm}
\usepackage[noend]{algpseudocode}
\usepackage[numbers,sort&compress]{natbib}
\usepackage{microtype}
\usepackage[font=small,labelfont=bf]{caption}
\usepackage{enumitem}
\usepackage{xcolor}
\usepackage{pifont}
\usepackage{tikz}
\usetikzlibrary{arrows.meta,positioning,fit,backgrounds}
\usepackage[colorlinks=true,linkcolor=blue!50!black,citecolor=blue!50!black,urlcolor=blue!50!black]{hyperref}

\setlist{nosep,leftmargin=1.4em}
\newtheorem{proposition}{Proposition}

\newcommand{\Sset}{\mathcal{S}}
\newcommand{\C}{\mathcal{C}}

\title{\textbf{Remembering a Number Without Revealing It:\\ Why Chat and Voice Assistants Cannot Host Bulls-and-Cows\\ and a Commit-Score-Reveal Fix}}
\author{\begin{tabular}{c@{\qquad}c}
Uygar Ege Kocakir & H. Fatih Ugurdag\\
\small\texttt{uygar.kocakir@ozu.edu.tr} & \small\texttt{fatih.ugurdag@ozyegin.edu.tr}
\end{tabular}\\[4pt]
\small Ozyegin University}
\date{}

\begin{document}
\maketitle

\begin{abstract}
Asked to ``come up with a five-digit number'' and score guesses in Bulls-and-Cows (Mastermind-style feedback such as $+2\,{-1}$), conversational assistants fail at a task that is trivial for a person. In five games in ChatGPT Voice, the assistant claimed to have chosen a number every time, contradicted its own answers in all five (first at turns 4-7), rejected a guess that was the only number still consistent with everything it had said, and at the end either named a number that contradicted most of its answers or said it had forgotten it. We argue that this is a structural property of today's deployment stack, not a reasoning slip: in a standard chat the only per-conversation state a model writes to is the transcript the user sees, the GPU key-value cache is a deterministic function of the model's input, and hidden reasoning is often discarded between turns. We formalise the game and compute exactly, over all $27{,}216$ secrets, how fast hosts without a secret contradict themselves: one that ignores the history does so in $79\%$ of games by turn~5, and one that respects it on $95\%$ of turns still does in $12\%$. A host that never contradicts itself is still not committed: it cannot be told apart from an honest one by outcomes, yet can stretch games by $41\%$. We propose \emph{Commit-Score-Reveal}: a sealed per-game secret outside the transcript, scoring in code, and a hash commitment the user verifies at the end, all implementable with today's function-calling APIs.
\end{abstract}

\section{Introduction}

Bulls-and-Cows is a code-breaking game. One player secretly chooses a number with five distinct digits, e.g.\ \texttt{12345}. The other player guesses, and receives feedback $(+a, -b)$: $a$ digits are correct and in the correct position, and $b$ digits occur in the secret but in a different position. Guessing \texttt{12385} against \texttt{12345} yields $+4$; guessing \texttt{45987} yields $-2$. The commercial board game Mastermind uses the same kind of feedback but allows repeated symbols: its classic version hides four pegs of six colours that may repeat, giving $6^4 = 1{,}296$ codes, and Knuth showed that the codebreaker can always win within five guesses \citep{knuth1976mastermind}. Throughout this paper, in the analysis and in the games played against ChatGPT, we use the traditional variant in which digits do not repeat. The game is trivial for a human host, because the host simply \emph{remembers} the number, and conventional hardware copes easily even with the harder role of the guesser: a population-based FPGA engine plays Mastermind faster than a high-end PC \citep{ugurdag2006mastermind}.

The motivating observation of this paper is that conversational assistants fail at exactly this trivial role. When one of the authors asked a speech-to-speech assistant (ChatGPT's voice mode) to pick a number and score guesses, the feedback soon became mutually inconsistent, and when asked which number it had chosen, the assistant replied that it had forgotten it. A small controlled measurement (\S\ref{sec:eval}) reproduces this in every game we played. The reason is easy to state. In a standard chat, the model remembers a conversation only through its context, and the part of the context it writes to is the conversation the user sees. A number it never wrote down is not there, and writing it into the chat would reveal it. A hidden part of the context, such as reasoning kept between turns or a tool, could hold the number, but only where the system provides one and the model actually uses it (\S\ref{sec:stack}). This is not an isolated quirk: it is an instance of what Baldelli et al.\ recently formalised as \emph{Private State Interactive Tasks} (PSITs), where an agent must keep hidden information while giving public answers consistent with it; they prove that an agent restricted to the public conversation history cannot both keep the secret unresolved and stay consistent with a fixed hidden state \citep{baldelli2026hangman}. Earlier, Shanahan et al.\ used the game of 20 Questions to argue that a dialogue agent is better described as a \emph{superposition} of characters than as a single character who has made up its mind \citep{shanahan2023roleplay}.

This short paper makes the problem concrete for a numeric game that is popular, easy to verify, and easy to simulate exactly. Our contributions are:
\begin{enumerate}
  \item A \textbf{systems-level account} of where conversational state physically lives in the inference stack (weights, transcript, KV cache, reasoning tokens, long-term memory, voice sessions), and why none of these can hold a private, persistent secret (\S\ref{sec:stack}).
  \item A \textbf{formalisation} of Bulls-and-Cows hosting, a specialisation of the PSIT impossibility result, and the observation that the best transcript-only host is \emph{outcome-equivalent} to an honest one but not committed (\S\ref{sec:formal}).
  \item An \textbf{exact simulation} over the full secret space quantifying how fast imperfect transcript-only hosts contradict themselves, and how much an adversarial non-committed host can stretch the game (\S\ref{sec:sim}).
  \item A minimal \textbf{Commit-Score-Reveal} (CSR) protocol that fixes the problem with one sealed state slot, one deterministic tool and a cryptographic commitment (\S\ref{sec:fix}).
  \item A \textbf{measurement in ChatGPT Voice}: five games under a fixed protocol, with guesses chosen by the solver, every answer checked against the exact consistent set, and public transcripts (\S\ref{sec:eval}).
\end{enumerate}

\subsection{How this work differs from Baldelli et al.}
\label{sec:diff}
The closest prior work is \citet{baldelli2026hangman}, and we build directly on their impossibility theorem. The two papers ask different questions. They ask \emph{whether} a language agent needs private memory, and answer it with model-managed memory architectures evaluated on Hangman and a diagnosis simulator. We ask \emph{where} the missing state should live in a deployed assistant, \emph{how quickly} its absence becomes visible, and \emph{how a user can verify} that a secret existed at all. Table~\ref{tab:diff} summarises the differences.

\begin{table}[h]
\centering\small
\caption{Scope of \citet{baldelli2026hangman} compared with this paper.}
\label{tab:diff}
\begin{tabularx}{\linewidth}{@{}>{\raggedright\arraybackslash}p{3.1cm}>{\raggedright\arraybackslash}X>{\raggedright\arraybackslash}X@{}}
\toprule
 & \citet{baldelli2026hangman} & This paper \\
\midrule
Task & Hangman; diagnosis simulator (DDXPlus) & Bulls-and-Cows (Mastermind-style numeric game) \\
Feedback the host computes & letter positions / yes--no answers & a two-number score that requires comparing all five digits \\
Who chooses the secret & the LLM (``privately choose'' a word) & a cryptographic RNG outside the model \\
Proposed fix & model-written private memory (agentic or workflow) & sealed write-once slot, scoring in code, hash commitment \\
Consistency of the fix & empirical: up to 100\% on Hangman, 16--20\% on diagnosis for one model & by construction, independent of model behaviour \\
User verifiability & not addressed & commit--reveal: user checks $\mathrm{SHA256}(\rho\|s^\star)=h$ \\
Measurement & fork test at a fixed turn ($t=4$) & exact consistent set after \emph{every} turn; first-contradiction turn \\
Non-committed hosts & not analysed & outcome-equivalence of the superposition host; adversarial host lengthens games by 41\% \\
Systems view & notes that reasoning traces are discarded & weights, KV/prefix cache, reasoning tokens, long-term memory, voice sessions \\
Models & open-weight text models (GPT-OSS, Qwen3); two-prompt check of chat products & ChatGPT Voice (GPT-Live): five games with public transcripts (\S\ref{sec:eval}) \\
\bottomrule
\end{tabularx}
\end{table}

Two differences matter most. In Bulls-and-Cows the \emph{scoring} step is itself error-prone, so CSR moves scoring into code instead of only adding memory; and the secret space is small enough to compute the exact consistent set after every answer, so a contradiction is located at the turn where it occurs. Their theorem explains why transcript-only agents fail; this paper measures how the failure unfolds and proposes a fix whose correctness does not depend on the model.

\section{Where does an assistant's ``memory'' live?}
\label{sec:stack}

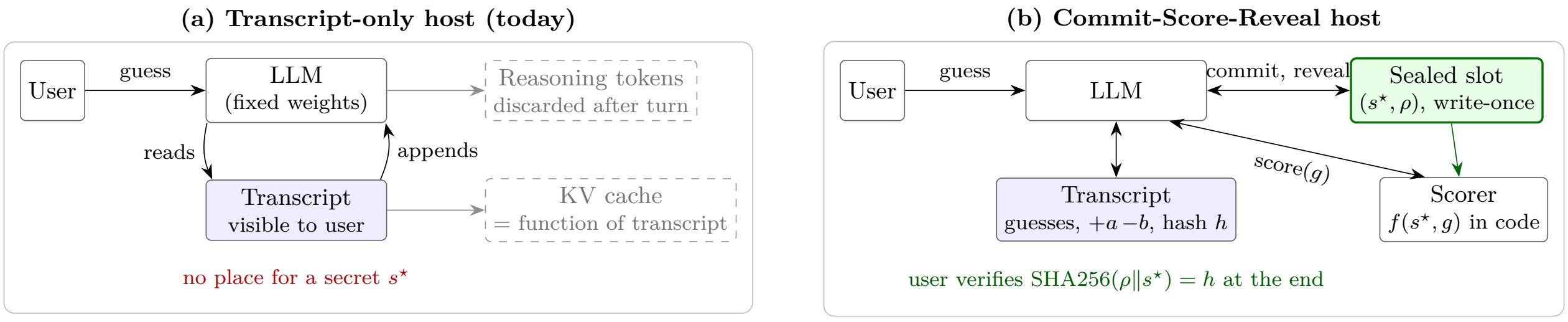
\begin{figure}[t]
\centering
\resizebox{\linewidth}{!}{%
\begin{tikzpicture}[
  font=\small, >={Stealth[length=2mm]},
  box/.style={draw=black!60, rounded corners=2pt, align=center, minimum height=8mm, inner sep=3pt, fill=white},
  pub/.style={box, fill=blue!7},
  gone/.style={box, dashed, text=black!55, draw=black!40},
  sealed/.style={box, fill=green!10, draw=green!45!black, thick},
  lab/.style={font=\footnotesize, align=center}]
\node[box] (u1) {User};
\node[box, right=1.6cm of u1, minimum width=2.4cm] (m1) {LLM\\[-1pt]\footnotesize (fixed weights)};
\node[pub, below=0.75cm of m1, minimum width=2.4cm] (t1) {Transcript\\[-1pt]\footnotesize visible to user};
\node[gone, right=1.3cm of m1] (r1) {Reasoning tokens\\[-1pt]\footnotesize discarded after turn};
\node[gone, right=1.3cm of t1] (k1) {KV cache\\[-1pt]\footnotesize $=$ function of transcript};
\draw[->] (u1) -- node[lab, above]{guess} (m1);
\draw[->] (m1.200) to[bend right=18] node[lab, left]{reads} (t1.160);
\draw[->] (t1.20) to[bend right=18] node[lab, right]{appends} (m1.340);
\draw[->, black!45] (m1) -- (r1);
\draw[->, black!45] (t1) -- (k1);
\node[lab, text=red!70!black, below=0.25cm of t1] (x1) {no place for a secret $s^\star$};
\begin{scope}[on background layer]
\node[draw=black!25, rounded corners, fit=(u1)(m1)(r1)(k1)(x1), inner sep=6pt, label={[font=\small\bfseries]above:(a) Transcript-only host (today)}] {};
\end{scope}
\begin{scope}[xshift=10.9cm]
\node[box] (u2) {User};
\node[box, right=1.6cm of u2, minimum width=2.4cm] (m2) {LLM};
\node[pub, below=0.75cm of m2, minimum width=2.4cm] (t2) {Transcript\\[-1pt]\footnotesize guesses, $+a\,{-b}$, hash $h$};
\node[sealed, right=1.9cm of m2] (s2) {Sealed slot\\[-1pt]\footnotesize $(s^\star,\rho)$, write-once};
\node[box, right=1.9cm of t2] (f2) {Scorer\\[-1pt]\footnotesize $f(s^\star, g)$ in code};
\draw[->] (u2) -- node[lab, above]{guess} (m2);
\draw[<->] (m2) -- node[lab, above]{commit, reveal} (s2);
\draw[<->] (m2.330) -- node[lab, sloped, below]{score$(g)$} (f2.140);
\draw[->, green!40!black] (s2) -- (f2);
\draw[<->] (m2) -- (t2);
\node[lab, text=green!35!black, below=0.25cm of t2] (x2) {user verifies $\mathrm{SHA256}(\rho\|s^\star)=h$ at the end};
\begin{scope}[on background layer]
\node[draw=black!25, rounded corners, fit=(u2)(m2)(s2)(f2)(x2), inner sep=6pt, label={[font=\small\bfseries]above:(b) Commit-Score-Reveal host}] {};
\end{scope}
\end{scope}
\end{tikzpicture}}
\caption{(a) In today's stack the only per-conversation state is the visible transcript; the KV cache is derived from it and hidden reasoning is dropped after each turn. (b) CSR adds a sealed, write-once slot and a deterministic scorer outside the model; only the hash $h$ and the feedback enter the transcript.}
\label{fig:arch}
\end{figure}

A useful way to see the problem is to ask, for each place that holds state during inference, whether it could store a secret number for the duration of a game without showing it to the user. Table~\ref{tab:state} summarises the answer and Figure~\ref{fig:arch}a sketches the resulting information flow.

\paragraph{Model weights.} Weights are shared across all users and are read-only at inference time; they cannot hold per-conversation information.

\paragraph{The transcript.} An LLM is a conditional distribution over the next token given the previous tokens \citep{vaswani2017attention,shanahan2023roleplay}. A chat ``conversation'' is implemented by re-feeding the accumulated messages at every turn: in Anthropic's description, each user message and assistant response accumulates in the context window and previous turns are preserved completely \citep{anthropic_context}. Apart from fixed instructions that the user does not see and, in some systems, reasoning kept between turns (below), the transcript is therefore the \emph{only} per-conversation state the model conditions on, and it is visible to the user. Writing the secret into it reveals it.

\paragraph{The KV cache (the hardware view).} During generation, transformer layers store attention keys and values for every processed token in accelerator memory. In serving systems such as vLLM, this KV cache is allocated and freed per request and occupies a large fraction of GPU memory (over 30\% on an A100 in the vLLM paper) \citep{kwon2023pagedattention}. Providers also keep \emph{prefix caches} so that a repeated prompt prefix need not be recomputed. These are opportunistic: cache hits require an exact prefix match, and caching works only if two requests land on the same machine \citep{openai_prompt_caching}. More fundamentally, the keys and values at each position are a deterministic function of the token prefix and the fixed weights. Sampling randomness enters the computation only through the token that is actually emitted. Hence \textbf{the KV cache contains no information beyond the tokens it was computed from}: if the model ``decides'' on a number without writing it as tokens, that decision leaves no trace anywhere in the hardware.

\paragraph{Hidden reasoning tokens.} Reasoning models generate private chain-of-thought tokens before answering, which might seem like a place to write the number. In many deployments these tokens do not survive the turn. OpenAI's guidance describes multi-turn conversations in which reasoning tokens are discarded after each turn while input and output tokens are carried forward \citep{openai_reasoning_items}, and on earlier Claude models the API strips previous thinking blocks from the history \citep{anthropic_context}; \citet{baldelli2026hangman} make the same observation for their agents. Newer interfaces can keep reasoning: OpenAI lets developers replay reasoning items or opt in to rendering earlier reasoning into the next turn \citep{openai_conversation_state}, and newer Claude models keep previous thinking blocks by default \citep{anthropic_context}. Kept reasoning is a private place that could hold a secret. It corresponds to the ``private chain-of-thought'' agent of \citet{baldelli2026hangman}, which restores consistency but needs about ten times more context than an explicit memory. It helps only if the model actually writes the secret there, and whether a consumer voice product keeps reasoning between turns is not documented.

\paragraph{Long-term memory features.} Products such as ChatGPT's memory store ``saved memories'' and insights from chat history to personalise future conversations \citep{openai_memory}. These stores are designed for user facts, are retrieval-oriented, and are user-inspectable, so they are neither private nor game-scoped. \citet{baldelli2026hangman} found that retrieval-based memory baselines fail their hidden-state consistency test just like vanilla models.

\paragraph{Voice mode.} Speech-to-speech assistants add latency pressure but not a new kind of state. ChatGPT Voice is currently powered by GPT-Live, a full-duplex model that listens and speaks at the same time \citep{openai_voice_help,openai_gptlive}. In its API, a GPT-Live session starts from prior text messages and receives later context through append events, beside the streamed audio and transcripts \citep{openai_live_sessions}; OpenAI's turn-based Realtime API likewise represents a session as a conversation of user and model items \citep{openai_realtime}. In both, unspoken context exists only if the application supplies it (for example through an append event or a tool); the default session gives the model no place of its own to keep a secret. A spoken game also moves the user's attention away from the written transcript and adds speech-recognition errors on digit strings, which make inconsistencies harder to notice and easier to cause.

\begin{table}[t]
\centering\small
\caption{Candidate locations for a hidden game secret in a deployed assistant.}
\label{tab:state}
\begin{tabularx}{\linewidth}{@{}>{\raggedright\arraybackslash}p{2.6cm}>{\raggedright\arraybackslash}p{3.4cm}>{\raggedright\arraybackslash}p{3.6cm}>{\raggedright\arraybackslash}X@{}}
\toprule
Location & Physical home & Lifetime & Can it hold a private secret across turns? \\
\midrule
Weights & accelerator memory, shared & model release & No: read-only, not per user \\
Transcript & server storage + client & conversation & Only by revealing it to the user \\
KV / prefix cache & GPU HBM, per request & request; prefix cache opportunistic & No: a deterministic function of the transcript \\
Reasoning tokens & KV cache during the turn & one turn in many APIs; kept by some newer ones & Only where kept and only if the model writes the secret there \\
Long-term memory & server database & across chats & Not private, not game-scoped, retrieval-based \\
\textbf{Sealed slot (\S\ref{sec:fix})} & server DB / CPU RAM & one game & \textbf{Yes}, sealed until reveal \\
\bottomrule
\end{tabularx}
\end{table}

\section{Formalisation}
\label{sec:formal}

Let $\Sset$ be the set of five-digit numbers with distinct digits and a non-zero leading digit, so $|\Sset| = 9\cdot9\cdot8\cdot7\cdot6 = 27{,}216$. For a secret $s$ and guess $g$, the feedback is $f(s,g) = (a,b)$ with $a = |\{i : s_i = g_i\}|$ and $b = |\{\text{digits of } g\} \cap \{\text{digits of } s\}| - a$. After $t$ rounds the transcript is $H_t = ((g_1,r_1),\dots,(g_t,r_t))$, and the \emph{consistent set} is
\[
  \C_t = \{ s \in \Sset : f(s,g_i) = r_i \ \text{for all } i \le t \}, \qquad \C_0 = \Sset .
\]
A transcript is \emph{satisfiable} if $\C_t \neq \emptyset$. A \textbf{committed host} draws $s^\star$ once and answers $r_t = f(s^\star, g_t)$; its transcripts are always satisfiable and $s^\star \in \C_t$ for all $t$. A \textbf{transcript-only host} answers $r_t \sim \pi(\cdot \mid H_{t-1}, g_t)$; by \S\ref{sec:stack}, every deployed LLM without a tool or persisted reasoning is of this type.

\begin{proposition}[specialisation of \citet{baldelli2026hangman}]
Let $t$ be a round at which $\C_{t-1}$ contains two secrets $s_1,s_2$ with $f(s_1,g_t)\neq f(s_2,g_t)$. Then no transcript-only host is consistent with a fixed secret: its reply distribution at round $t$ is the same whether the ``intended'' secret is $s_1$ or $s_2$, so it can be correct for at most one of them, and which one is not determined before the reply is sampled.
\end{proposition}
\begin{proof}[Proof sketch]
The reply depends only on $(H_{t-1},g_t)$ and fresh sampling noise. Since both $s_1$ and $s_2$ are consistent with $H_{t-1}$, no variable available to $\pi$ distinguishes them; an earlier ``choice'' that was not written into $H_{t-1}$ is not available (\S\ref{sec:stack}).
\end{proof}

The proposition does not say that a transcript-only host must lose consistency. The best it can do is the \textbf{superposition host}: sample $r_t$ as the feedback of a uniformly random element of $\C_{t-1}$. Its transcripts are always satisfiable and, by the principle of deferred decisions, have exactly the same distribution as those of a committed host whose secret is uniform on $\Sset$. Such a host is \emph{outcome-equivalent} to an honest one, yet the ``secret'' it reveals at the end is decided only by the last answers. Two practical problems remain. First, to behave this way an LLM must re-solve a constraint-satisfaction problem over all previous guesses inside every reply, a deduction task on which current models struggle even when the hints are given explicitly \citep{golde2025mastermindeval}. Second, nothing prevents a non-committed host from choosing answers that keep $\C_t$ large, the numeric analogue of ``Evil Hangman'' \citep{barbay2020evilhangman}. We model these cases with four hosts:
\begin{itemize}
  \item \emph{memoryless}: answers each guess plausibly in isolation, sampling $r_t$ from the feedback distribution of $g_t$ over all of $\Sset$ (history ignored);
  \item \emph{mixed($p$)}: with probability $p$ behaves as the superposition host, otherwise as the memoryless host, a crude model of a system that tracks the history correctly on a fraction $p$ of turns;
  \item \emph{superposition}: mixed($1$);
  \item \emph{evil}: chooses the feedback class that keeps the largest consistent set.
\end{itemize}

\section{Simulation}
\label{sec:sim}

\paragraph{Setup.} We enumerate all $27{,}216$ secrets and score guesses exactly with vectorised integer arithmetic. The guesser is a sensible human-like solver that always guesses a random element of the current consistent set (if the set becomes empty, it guesses at random). We run 400 games for the consistent-set curves, 1{,}500 games for each value of $p$ in the satisfiability experiments, and 300 games per host for game length, with fixed seeds; the code is released with the paper.

\begin{figure}[t]
\centering
\includegraphics[width=\linewidth]{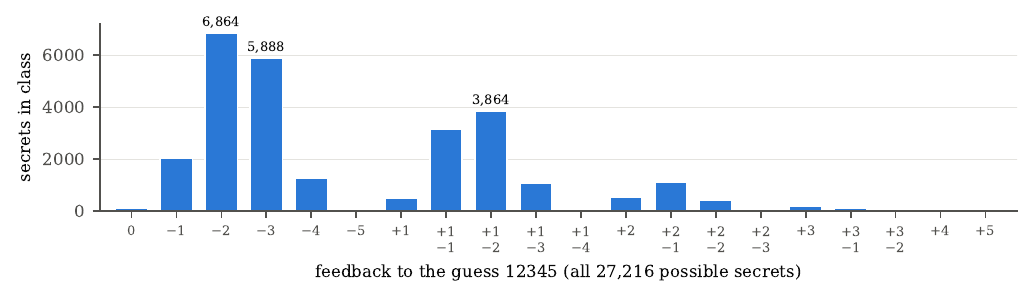}
\caption{Number of secrets that produce each feedback for the first guess \texttt{12345}. A host that has a secret gives one of these 20 answers deterministically; a host without one must pick an answer that remains possible given every earlier answer.}
\label{fig:classes}
\end{figure}

\begin{figure}[t]
\centering
\begin{tabular}{cc}
\includegraphics[width=0.47\linewidth]{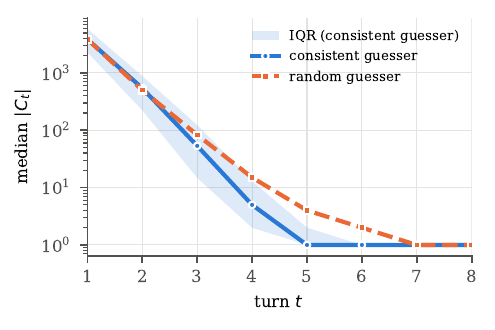} &
\includegraphics[width=0.47\linewidth]{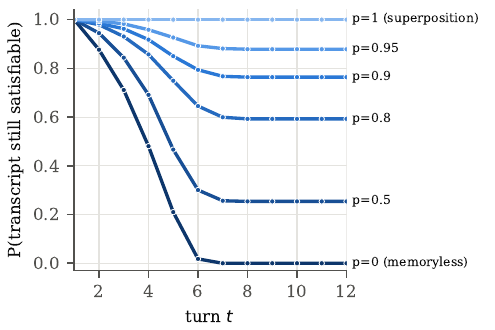}\\
\small (a) consistent-set size, committed host & \small (b) transcript satisfiability, transcript-only hosts
\end{tabular}
\caption{(a) Median number of secrets consistent with the transcript (log scale; shaded band: interquartile range). (b) Fraction of games whose transcript is still satisfiable at turn $t$; games end when the guess is correct. $p$ is the per-turn probability that the host's answer correctly respects the history.}
\label{fig:sim}
\end{figure}

\paragraph{How much the hidden secret constrains each answer.} Figure~\ref{fig:classes} shows how a single guess, \texttt{12345}, splits $\Sset$ into 20 feedback classes. The distribution is very uneven: the classes $-2$ and $-3$ together hold $47\%$ of all secrets, while $+3\,{-2}$ holds only ten. One answer therefore carries $3.06$ bits of information (out of a possible $\log_2 20 = 4.32$), and the right answer depends sharply on the secret. Figure~\ref{fig:sim}a shows the median size of $\C_t$ with an honest committed host. The median consistent set falls from $3{,}864$ after one guess to $542$, $53.5$ and $5$, and reaches a single secret by turn~5. In other words, by the middle of a typical game almost every possible answer to a new guess is already ruled out by the history, and a host without the secret must reproduce that history-dependent constraint from scratch in every reply.

\paragraph{How fast imperfect hosts contradict themselves.} Figure~\ref{fig:sim}b reports the probability that the transcript is still satisfiable. The memoryless host is contradictory in $12\%$ of games after two turns, $52\%$ after four, $79\%$ after five, $98\%$ after six and $100\%$ by turn~7. This matches the user experience that motivated this work: the first one or two answers look reasonable, and the game then collapses. Partial tracking helps but does not solve the problem: even with $p = 0.95$, i.e.\ a host that respects the history on 19 of 20 turns, $12\%$ of games contain a contradiction; with $p=0.9$ the figure is $24\%$ and with $p=0.8$ it is $41\%$. Because errors are absorbing (an empty $\C_t$ stays empty), per-turn reliability compounds over the game, and the contradiction rate falls only slowly as $p$ approaches one.

\paragraph{Non-commitment is invisible in outcomes.} Table~\ref{tab:len} compares game lengths. The superposition host is statistically indistinguishable from an honest committed host (both $5.92$ guesses on average, with nearly identical spread), as the deferred-decision argument predicts. A user therefore cannot detect a non-committed host from how games turn out. The evil host, which is also always consistent, lengthens games by $41\%$ on average (to $8.37$ guesses); only 2 of its 300 games ended before the eighth guess. Consistency checks alone cannot rule this out; only a commitment made before the first guess can.

\begin{table}[t]
\centering\small
\caption{Number of guesses needed by the consistent-set guesser (300 games per host). All three hosts produce satisfiable transcripts.}
\label{tab:len}
\begin{tabular}{@{}lccccc@{}}
\toprule
Host & mean & std & min & 90th pct. & max \\
\midrule
Committed (honest) & 5.92 & 0.98 & 3 & 7 & 8 \\
Superposition (transcript-only, best case) & 5.92 & 0.98 & 2 & 7 & 8 \\
Evil (transcript-only, adversarial) & 8.37 & 0.53 & 7 & 9 & 10 \\
\bottomrule
\end{tabular}
\end{table}

\section{A fix: Commit-Score-Reveal}
\label{sec:fix}

The analysis suggests three separate requirements: the secret must \emph{exist} outside the transcript, the feedback must be \emph{computed} rather than recalled, and the user must be able to \emph{verify} afterwards that the secret was fixed from the start. Algorithm~\ref{alg:csr} and Figure~\ref{fig:arch} meet all three with components that current chat and realtime voice APIs already support through function calling.

\begin{algorithm}[t]
\caption{Commit-Score-Reveal (CSR) game hosting}
\label{alg:csr}
\begin{algorithmic}[1]
\State \textbf{commit}(\textit{conv\_id}): server draws $s^\star \in \Sset$ and a 128-bit salt $\rho$ from a cryptographic RNG
\State \quad store $(s^\star,\rho)$ in a \emph{sealed slot} keyed by (\textit{conv\_id}, \textit{game\_id}); return \textit{game\_id} and $h = \mathrm{SHA256}(\rho \,\|\, s^\star)$
\State \quad the model announces: ``I have locked a number; fingerprint \texttt{h[0:8]}''
\For{each user guess}
  \State the model normalises the guess and, in voice mode, reads the digits back for confirmation
  \State \textbf{score}(\textit{game\_id}, $g$): server validates $g$ and returns $f(s^\star, g)$; the model only verbalises ``$+a\,{-b}$''
\EndFor
\State \textbf{reveal}(\textit{game\_id}) on success or surrender: return $(s^\star,\rho)$; the client checks $\mathrm{SHA256}(\rho\|s^\star) = h$; slot is deleted
\end{algorithmic}
\end{algorithm}

\paragraph{Sealed slot.} The slot is a small per-game record in ordinary server memory or a database, not in the accelerator, so it is unaffected by KV-cache eviction, by requests being routed to different machines, and by context truncation or compaction. It is write-once; the model has no tool that reads $s^\star$ before \textbf{reveal}, so a user cannot extract the number by asking (``what number did you pick?''), and the slot is excluded from long-term memory. Drawing $s^\star$ with a system RNG rather than letting the model ``pick'' also avoids the strong mode collapse that \citet{baldelli2026hangman} report in model-chosen secrets.

\paragraph{Deterministic scoring.} Moving $f$ into code removes the need for the model to compare digits in its head; current models already struggle to combine Mastermind feedback correctly \citep{golde2025mastermindeval}. Scoring is $O(1)$; the added latency is one tool round-trip per guess.

\paragraph{Commitment.} Publishing $h$ before the first guess is a standard commit--reveal scheme \citep{blum1983coin}: the hash hides $s^\star$ (the salt prevents a brute-force search over only $27{,}216$ candidates), while the binding property means the host cannot later reveal a different number. This makes the evil host of Table~\ref{tab:len} detectable. In a voice interface, the first few hexadecimal characters can be spoken and the full hash shown in the app.

\paragraph{Lighter alternatives.} A cheaper variant inserts the secret as a hidden system message that the client never displays. It solves consistency, since the secret is now in the context, but it still relies on in-head scoring, can leak the number when the model is asked for it, and offers no verifiability. Opt-in persisted reasoning \citep{openai_conversation_state} behaves similarly at a higher token cost. When neither tools nor hidden state are available, the honest behaviour is for the assistant to say that it cannot hold a secret, and to offer to play as the guesser instead.

\section{Measurement in ChatGPT Voice}
\label{sec:eval}

\paragraph{Protocol.} On 26 November 2026 we played five games against ChatGPT Voice, which is powered by GPT-Live \citep{openai_voice_help}, on a ChatGPT Plus account, in Turkish. Each game used a new chat, and ChatGPT's memory feature was left on, as set on the account. ChatGPT ran on a phone placed next to a laptop. A referee web page on the laptop read the opening request and every guess aloud with the operating system's Turkish text-to-speech voice, so ChatGPT received spoken input exactly as from a user. The experimenter entered each spoken answer $(+a,-b)$ on the page, which ran the exact solver of \S\ref{sec:sim} after every answer. Guesses were not chosen by the experimenter: the page drew each one uniformly from the current consistent set $\C_{t-1}$, the same guesser as in the simulation. After the first contradiction the game continued for two more turns, then the page asked ``I give up. What was your number? Can you say it digit by digit?'' The opening request, the guesses and the final question were identical in form across games. The shared transcript of every game is public (Appendix~\ref{app:links}), so each recorded answer can be checked by any reader.

\begin{table}[t]
\centering\small
\caption{The five ChatGPT Voice games. $|\C_t|$ after each answer, shown up to the first contradiction (\textbf{0}); the set stays empty for the remaining turns. ``Left before'' is $|\C_{t-1}|$ at that turn; ``consistent answers'' are the answers that at least one remaining secret would produce for the guess.}
\label{tab:chatgpt}
\begin{tabularx}{\linewidth}{@{}c>{\raggedright\arraybackslash}Xccl>{\raggedright\arraybackslash}p{3.6cm}@{}}
\toprule
Game & $|\C_t|$ per turn & Left before & Said & Consistent answers & Number named at the end \\
\midrule
1 & 3{,}864, 186, 3, \textbf{0} & 3 & $+1\,{-3}$ & $+5$, $+2\,{-3}$, $+1\,{-2}$ & \texttt{12450}: contradicts 8 of 8 answers \\
2 & 3{,}864, 186, 18, 2, \textbf{0} & 2 & $+1\,{-2}$ & $+5$, $+3\,{-2}$ & \texttt{46780}: contradicts 5 of 6 \\
3 & 7{,}020, 1{,}082, 220, 30, \textbf{0} & 30 & $+1\,{-3}$ & ten answers, incl.\ $+5$ & \texttt{63751}: contradicts 4 of 5 \\
4 & 6{,}864, 810, 167, 20, 2, 1, \textbf{0} & 1 & $+2$ & $+5$ only & \texttt{34785}: contradicts 5 of 7 \\
5 & 1{,}272, 159, 20, 5, 2, \textbf{0} & 2 & $+1\,{-3}$ & $+5$, $+1\,{-4}$ & said it had forgotten the number \\
\bottomrule
\end{tabularx}
\end{table}

\begin{figure}[t]
\centering
\includegraphics[width=\linewidth]{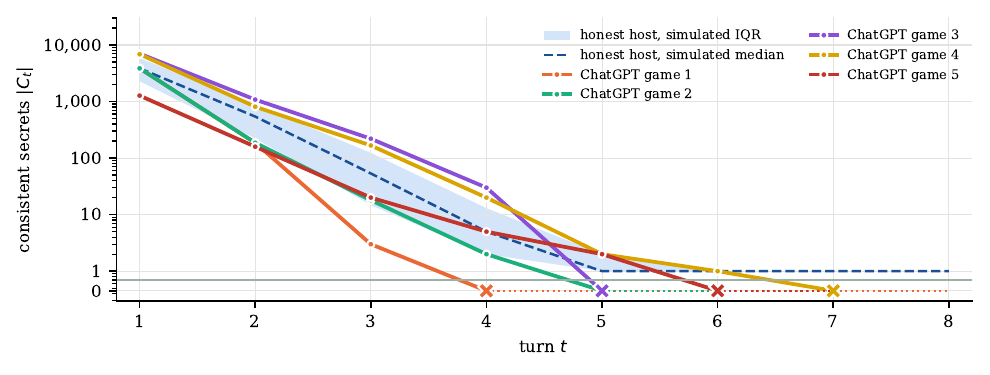}
\caption{Secrets consistent with ChatGPT's answers in the five games (solid lines) over the simulated honest host of \S\ref{sec:sim} (band: interquartile range of 400 games). Until the first contradiction (\ding{54}) ChatGPT's answers narrow the set the way an honest host's would; then no five-digit number fits them any more.}
\label{fig:chatgpt}
\end{figure}

\paragraph{Every game contradicted itself.} Table~\ref{tab:chatgpt} and Figure~\ref{fig:chatgpt} summarise the games (33 scored turns; median 9.4\,s between turns). ChatGPT said that it had picked a number at the start of all five games, and its answers in all five became unsatisfiable, first at turns 4, 5, 5, 7 and 6. Until then its answers were not random: the consistent set shrank at the same rate as with an honest host, so each early answer, and the early answers jointly, looked exactly like play with a real secret. The experimenter recorded no misheard guesses. Memory being on did not help: consistent with \S\ref{sec:stack}, a long-term store designed for user facts is not used as a game's hidden state.

\paragraph{It rejects the number its own answers point to.} Because the guesser always picks a remaining candidate, every guess was a number that matched all of ChatGPT's previous answers, and ``$+5$, you win'' was always a consistent reply. ChatGPT never gave it. In four games the contradiction came when at most three candidates were left. In game~4 its first six answers left exactly one number, \texttt{73685}; we guessed it, and the only answer consistent with its own history was $+5$, but ChatGPT said $+2$. In game~3, thirty candidates admitted ten different consistent answers and ChatGPT gave an eleventh. In game~5 it gave the same answer, $+1\,{-3}$, five times in a row (turns 2--6), suggesting that it repeated a recent answer rather than scored the guess.

\paragraph{The number at the end.} Asked for its number, ChatGPT named a valid five-digit number in four games. None of them fits its own answers: each contradicts between four and all eight of the answers it had given (Table~\ref{tab:chatgpt}). In game~5 it said that it had forgotten the number. These are the two behaviours predicted for a transcript-only host: a ``secret'' produced at reveal time, or an admission that there was none.

\paragraph{Comparison with the simulated hosts.} We asked which per-turn reliability $p$ of the mixed($p$) host of \S\ref{sec:formal} best explains the observed first-contradiction turns, simulating 2{,}000 games for each $p$ on a grid. The likelihood is highest near $p=0.3$, and $p \le 0.6$ lies within the approximate 95\% likelihood interval. Under the mixed host, all five games contradicting themselves has probability $0.013$ at $p=0.8$ and $0.0005$ at $p=0.9$. Within this idealised model, ChatGPT Voice respected its own history on well under 80\% of turns and was far from the always-consistent superposition host. With five games this is an order-of-magnitude statement, not a precise estimate.

\section{Discussion and limitations}

The failure analysed here is broader than one game. Twenty Questions, Hangman, Battleship, card games, quizzes with a hidden answer, and negotiations with a private reservation price all require an agent to hold information it must not say. Recent benchmarks already include such settings: \citet{badola2025multiturn} use a Twenty Questions task to test logical consistency over a dialogue, and \citet{marioriyad2026lying} clone a 20 Questions game into parallel worlds to test whether models deny their chosen object. Memory architectures such as MemGPT \citep{packer2023memgpt} move information between tiers but are designed for recall, not secrecy. There is also an honesty dimension: an assistant that says ``I have picked a number'' when no number exists anywhere in the system is making a false statement about its own state, even if unintentionally. In our games ChatGPT claimed a number five times out of five and, when asked for it, invented one four times. Admitting that it has ``forgotten'' the number, as it did once, is more honest than inventing one, but it comes too late: by then the user has spent the whole game scoring guesses against a secret that never existed. The honest point to say so is the beginning, before the first guess.

Our study has clear limitations. The measurement in \S\ref{sec:eval} is a small case study: five games, one account, one day and one language, with answers entered by hand by the experimenter (the public transcripts allow every entry to be checked) and ChatGPT's memory left on, so later games could in principle draw on earlier ones even though each was played in a new chat. A larger study, for example through the GPT-Live API, could add a fork test in the style of \citet{baldelli2026hangman} and compare the same protocol with CSR enabled. The simulated hosts are idealised models of LLM behaviour, and the internal architecture of commercial voice assistants is not public; \S\ref{sec:stack} relies on published API documentation as the closest available description. The mixed($p$) host assumes independent per-turn errors, whereas real models may fail in correlated ways (for example, more often late in long games). Finally, CSR fixes the problem by \emph{adding} state outside the model; whether models can be trained to maintain compact private state natively remains an open question.

\section{Conclusion}

A chat or voice assistant cannot keep a number in mind because, in the standard deployment stack, there is no mind to keep it in: the visible transcript is the only per-conversation state the model writes to, the hardware cache is a function of the model's input, and private reasoning is often discarded between turns. Exact simulation of Bulls-and-Cows shows that hosts which answer without truly tracking the history contradict themselves within a few turns, and ChatGPT Voice did so in all five games we played, by turn~7 at the latest, and that even a perfectly consistent transcript-only host is not committed and cannot be distinguished from an honest one by outcomes. A sealed per-game slot, a deterministic scoring tool and a hash commitment are enough to turn the assistant into an honest, verifiable host, and they can be implemented today with existing function-calling interfaces.

\paragraph{Code and data availability.} The simulation scripts (\texttt{mastermind\_sim.py}, \texttt{figs2.py}), the analysis of the ChatGPT games (\texttt{analyze\_chatgpt.py}), the recorded games (\texttt{games.json}) and all raw results are provided as ancillary files.

\appendix
\section{Transcripts of the ChatGPT Voice games}
\label{app:links}
\begin{tabular}{@{}cl@{}}
Game 1 & \url{https://chatgpt.com/share/6b086e3f-c320-83eb-a808-9b7834393490}\\
Game 2 & \url{https://chatgpt.com/share/6b086e52-2870-83eb-b1f8-81d169251ceb}\\
Game 3 & \url{https://chatgpt.com/share/6b086e61-823c-83ed-9c32-a032a5f8ddaa}\\
Game 4 & \url{https://chatgpt.com/share/6b086e6f-a270-83ed-a453-216229d3bf54}\\
Game 5 & \url{https://chatgpt.com/share/6b086e7c-50f8-83eb-ba75-716e0f5d7b09}\\
\end{tabular}


\begin{thebibliography}{99}\small\sloppy

\bibitem[Badola et~al.(2025)]{badola2025multiturn}
K.~Badola, J.~Simon, A.~Hosseini, S.~M. Mc~Carthy, T.~Munkhdalai, A.~Goyal, T.~Ko\v{c}isk\'{y}, S.~Upadhyay, B.~Fatemi, and M.~Kazemi.
\newblock Multi-turn puzzles: Evaluating interactive reasoning and strategic dialogue in {LLMs}.
\newblock arXiv:2508.10142, 2025.

\bibitem[Baldelli et~al.(2026)]{baldelli2026hangman}
D.~Baldelli, A.~Parviz, A.~Zouaq, and S.~Chandar.
\newblock {LLMs} can't play {H}angman: On the necessity of a private working memory for language agents.
\newblock In \emph{Conference on Lifelong Learning Agents (CoLLAs)}, 2026 (accepted). arXiv:2601.06973.

\bibitem[Barbay and Subercaseaux(2020)]{barbay2020evilhangman}
J.~Barbay and B.~Subercaseaux.
\newblock The computational complexity of {E}vil {H}angman.
\newblock arXiv:2003.10000, 2020.

\bibitem[Blum(1983)]{blum1983coin}
M.~Blum.
\newblock Coin flipping by telephone: A protocol for solving impossible problems.
\newblock \emph{ACM SIGACT News}, 15(1):23--27, 1983. doi:10.1145/1008908.1008911.

\bibitem[Golde et~al.(2025)]{golde2025mastermindeval}
J.~Golde, P.~Haller, F.~Barth, and A.~Akbik.
\newblock {MastermindEval}: A simple but scalable reasoning benchmark.
\newblock In \emph{ICLR 2025 Workshop on Reasoning and Planning for Large Language Models}, 2025. arXiv:2503.05891.

\bibitem[Knuth(1976)]{knuth1976mastermind}
D.~E. Knuth.
\newblock The computer as master mind.
\newblock \emph{Journal of Recreational Mathematics}, 9(1):1--6, 1976.

\bibitem[Kocak et~al.(2006)]{ugurdag2006mastermind}
Y.~S. Kocak, S.~Goren, S.~Dedeoglu, O.~Baskirt, Y.~Sahin, and H.~F. Ugurdag.
\newblock Population-Based {FPGA} Solution to {Mastermind} Game.
\newblock In \emph{First NASA/ESA Conference on Adaptive Hardware and Systems (AHS'06)}, pages 237--246. IEEE, June 2006. doi:10.1109/AHS.2006.57.

\bibitem[Kwon et~al.(2023)]{kwon2023pagedattention}
W.~Kwon, Z.~Li, S.~Zhuang, Y.~Sheng, L.~Zheng, C.~H. Yu, J.~E. Gonzalez, H.~Zhang, and I.~Stoica.
\newblock Efficient memory management for large language model serving with {PagedAttention}.
\newblock In \emph{Proc.\ ACM SOSP}, 2023. arXiv:2309.06180.

\bibitem[Marioriyad et~al.(2026)]{marioriyad2026lying}
A.~Marioriyad, A.~Nouri, M.~H. Rohban, and M.~Soleymani~Baghshah.
\newblock Lying to win: Assessing {LLM} deception through human-{AI} games and parallel-world probing.
\newblock arXiv:2603.07202, 2026.

\bibitem[OpenAI(a)]{openai_reasoning_items}
OpenAI.
\newblock Better performance from reasoning models using the {Responses API}.
\newblock OpenAI Cookbook. \url{https://developers.openai.com/cookbook/examples/responses_api/reasoning_items}. Accessed Nov.\ 2026.

\bibitem[OpenAI(b)]{openai_conversation_state}
OpenAI.
\newblock Conversation state.
\newblock API documentation. \url{https://developers.openai.com/api/docs/guides/conversation-state}. Accessed Nov.\ 2026.

\bibitem[OpenAI(c)]{openai_prompt_caching}
OpenAI.
\newblock Prompt caching 201.
\newblock OpenAI Cookbook. \url{https://developers.openai.com/cookbook/examples/prompt_caching_201}. Accessed Nov.\ 2026.

\bibitem[OpenAI(d)]{openai_realtime}
OpenAI.
\newblock Realtime conversations.
\newblock API documentation. \url{https://developers.openai.com/api/docs/guides/realtime-conversations}. Accessed Nov.\ 2026.

\bibitem[OpenAI(e)]{openai_memory}
OpenAI.
\newblock Memory and new controls for {ChatGPT}.
\newblock \url{https://openai.com/index/memory-and-new-controls-for-chatgpt/}, Feb.\ 2024 (updated June 2025).

\bibitem[OpenAI(f)]{openai_voice_help}
OpenAI.
\newblock {ChatGPT Voice}.
\newblock Help Center. \url{https://help.openai.com/en/articles/8400625-voice-mode}. Accessed Nov.\ 2026.

\bibitem[OpenAI(g)]{openai_gptlive}
OpenAI.
\newblock Build more natural voice experiences with {GPT-Live-1} in the {API}.
\newblock \url{https://openai.com/index/introducing-gpt-live-1-in-the-api/}, 10 Sept.\ 2026.

\bibitem[OpenAI(h)]{openai_live_sessions}
OpenAI.
\newblock Managing {GPT-Live} sessions.
\newblock API documentation. \url{https://developers.openai.com/api/docs/guides/live-conversations}. Accessed Nov.\ 2026.

\bibitem[Anthropic()]{anthropic_context}
Anthropic.
\newblock Context windows.
\newblock Claude Platform documentation. \url{https://platform.claude.com/docs/en/build-with-claude/context-windows}. Accessed Nov.\ 2026.

\bibitem[Packer et~al.(2023)]{packer2023memgpt}
C.~Packer, S.~Wooders, K.~Lin, V.~Fang, S.~G. Patil, I.~Stoica, and J.~E. Gonzalez.
\newblock {MemGPT}: Towards {LLMs} as operating systems.
\newblock arXiv:2310.08560, 2023.

\bibitem[Shanahan et~al.(2023)]{shanahan2023roleplay}
M.~Shanahan, K.~McDonell, and L.~Reynolds.
\newblock Role play with large language models.
\newblock \emph{Nature}, 623(7987):493--498, 2023. doi:10.1038/s41586-023-06647-8.

\bibitem[Vaswani et~al.(2017)]{vaswani2017attention}
A.~Vaswani, N.~Shazeer, N.~Parmar, J.~Uszkoreit, L.~Jones, A.~N. Gomez, \L{}.~Kaiser, and I.~Polosukhin.
\newblock Attention is all you need.
\newblock In \emph{Advances in Neural Information Processing Systems 30}, 2017.

\end{thebibliography}
\end{document}